\documentstyle[fps,HEP93]{article}

\newcommand{\al}{\alpha}

\newcommand{\Sg}{\Sigma}
\newcommand{\dl}{\delta}
\newcommand{\ep}{\varepsilon}

\newcommand{\vep}{\varepsilon}

\newcommand{\kp}{\kappa}
\newcommand{\lm}{\lambda}

\newcommand{\vphibar}{\overline{\varphi}}

\newcommand{\sg}{\sigma}

\newcommand{\hmu}{\hat{\mu}}
\newcommand{\phat}{\hat{p}^2}

\newcommand{\half}{\mbox{{\small $\frac{1}{2}$}} }

\newcommand{\ra}{\rightarrow}

\newcommand{\be}{\begin{equation}}
\newcommand{\ee}{\end{equation}}
\newcommand{\bea}{\begin{eqnarray}}
\newcommand{\eea}{\end{eqnarray}}

\newcommand{\beq}{\begin{equation}}
\newcommand{\eeq}{\end{equation}}

\newcommand{\lb}{\label}

\def \3{\ss}

\begin{document}
\title{NON-PERTURBATIVE INVESTIGATION OF
THE FERMION-HIGGS SECTOR OF THE STANDARD MODEL}
\author{Wolfgang Bock$^1$, Christoph Frick$^2$, Jan Smit$^3$ and Jeroen C.
Vink$^1$ \\
 \it $^1$ University of California, San Diego,
     Department of Physics-0319,
     La Jolla, CA 92093-0319, USA \\
 \it $^2$ HLRZ c/o KFA J\"ulich, P.O. Box 1913, 5170 J\"ulich, Germany and \\
 \it Institute of Theoretical Physics E, RWTH
     Aachen, Sommerfeldstr., 5100 Aachen, Germany\\
 \it $^3$ Institute of Theoretical Physics, University of
     Amsterdam, Valckenierstraat 65,  \\
 \it 1018 XE Amsterdam,
     The Netherlands
 \
 }
\vspace{2cm}
\abstract{\rightskip=1.5pc
          \leftskip=1.5pc
We present results for the renormalized quartic
self-coupling $\lm_R$ and the
renormalized Yukawa coupling $y_R$ in a fermion-Higgs model
with two SU(2) doublets, indicating that with the  standard
lattice regulator these couplings cannot become
very strong.
}
%
%
\maketitle
\vskip 5mm
{\bf \noindent The Lattice Model}
\vskip 2.5mm
It is an important issue to investigate within a non-perturbative
regularization scheme whether the quartic self-coupling and Yukawa coupling
of the fermion-Higgs sector of the Standard model (StM) may become
strong when
increasing the bare couplings to very large values while keeping
the cut-off roughly constant. For this
it is desirable to construct a lattice fermion-Higgs model
with a fermion content which is as close as possible to that of
the StM. This is a non-trivial issue
since a naive transcription of the continuum la\-gran\-gian
with one SU(2) doublet
(with gauge interactions switched off for simplicity)
leads to the large  number of
16 doublets on the lattice because of the species doubling phenomenon.
Two strategies have been worked out so far
to reduce this large number of mass-degenerate SU(2) doublets to one:
The mirror fermion model [1]          and the reduced staggered fermion model
[2]. In this contribution we shall only        focus on
the second method. \\
Reduced staggered fermions are represented on the lattice by the one-component
real Grassmann field $\chi$ which describes in the continuum limit
two Dirac flavors. It is not immediately clear how to access these flavor
degrees of freedom since the staggered spin and flavor degrees of freedom
are spread out over the lattice.  By using however the method outlined
in ref.~[2]       it is possible to couple these
two staggered flavors as an isospin-doublet to the Higgs field.
For more details about the staggered method we refer the reader
to refs.~[2,3,4].
\begin{figure}[t,b]
\centerline{
\fpsxsize=5.5cm
\fpsbox{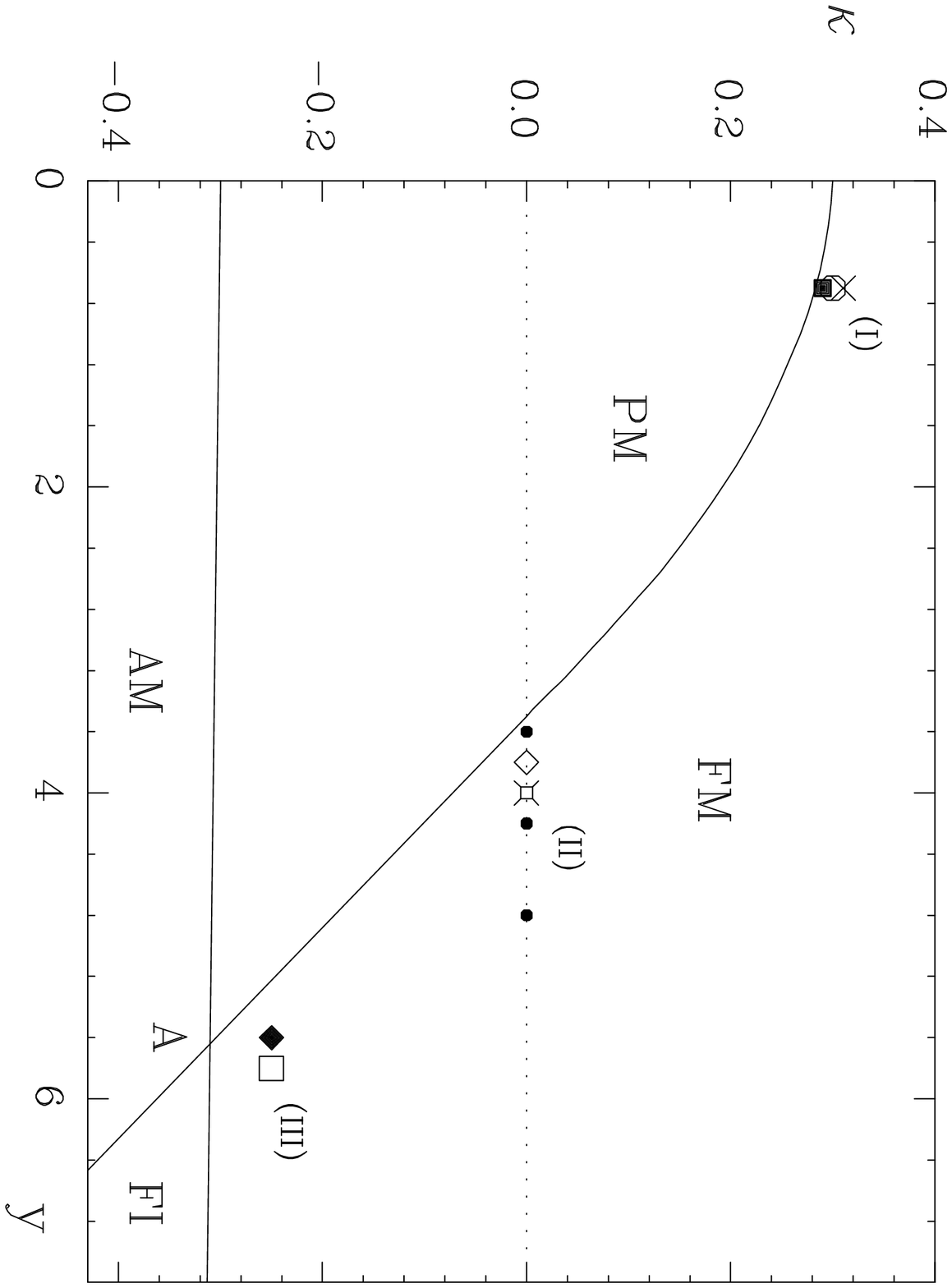}
}
\vspace*{-0.9cm}
\hspace*{0.6cm}
\centerline{
\fpsxsize=7.0cm
\fpsbox{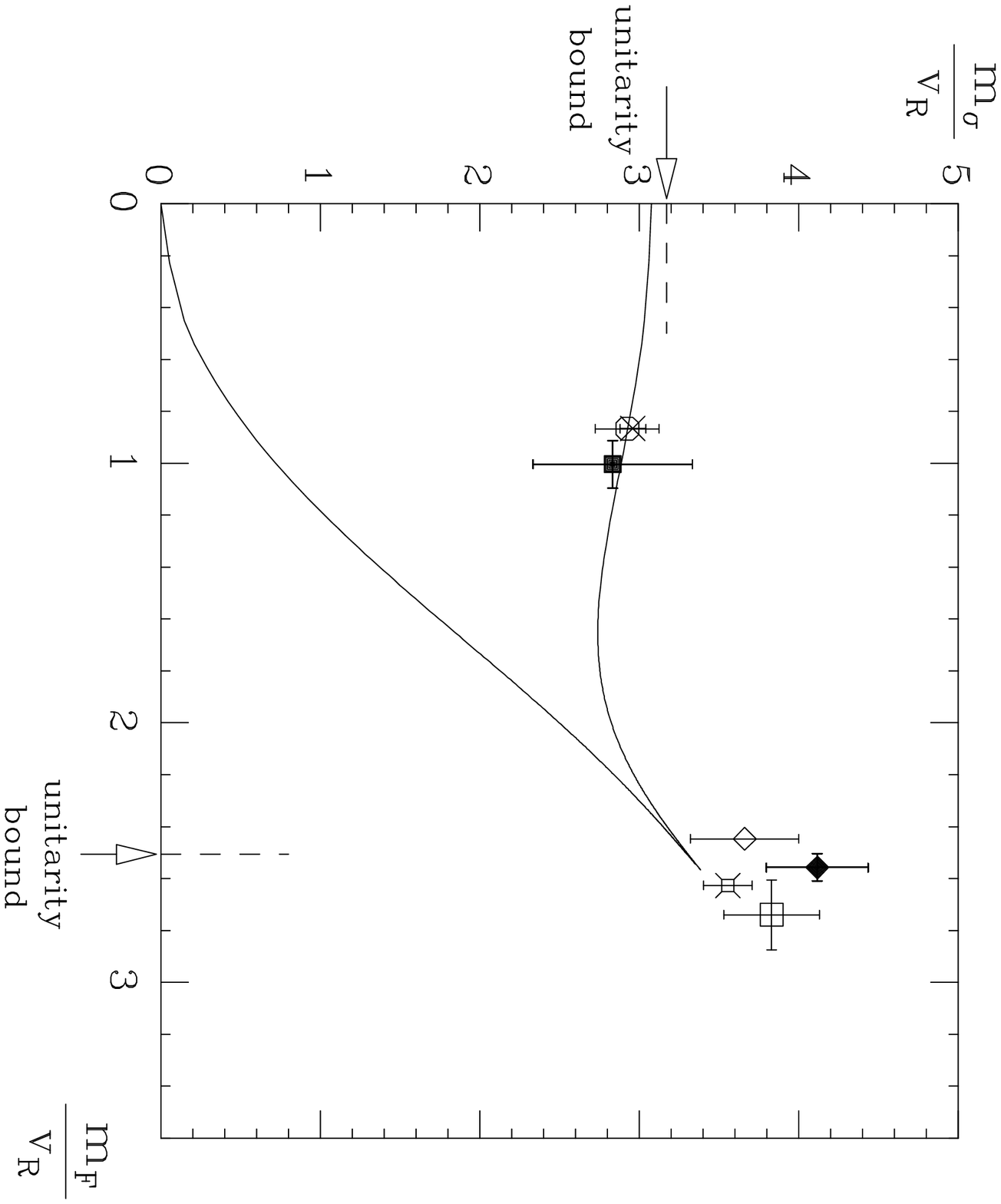}
}
\vspace*{-1.1cm}
\caption{Upper figure: Phase diagram at $\lm=\infty$.
Lower figure: $m_{\sg}/v_R$ as a function of $m_F/v_R$.
}
\label{fig:1}
\end{figure}
The final form of the one-doublet action in terms
of the  staggered field $\chi$ reads
\bea
 S_F = -\half  \sum_{x \mu} \chi_x \chi_{x+\hmu}
       ( \eta_{\mu x} + y \ep_x \zeta_{\mu x} \vphibar_{\mu x} )
       \;, \lb{SCHI}
\eea
where $\vphibar_{\mu  x} = \frac{1}{16} \sum_b \varphi_{\mu, x-b}$
is the average of the scalar field over a lattice hypercube and
$\eta_{\mu x}=(-1)^{x_1+\cdots +x_{\mu-1}}$,
$\zeta_{\mu x}=(-1)^{x_{\mu+1}+\cdots +x_4}$ are the usual staggered
sign factors. The complete
action is given by
$S=S_F+S_H$, where
$S_H= \sum_{x}[2\kp \sum_{\mu} \varphi_{\al x} \varphi_{\al,x+\hmu} -
\varphi_{\al x} \varphi_{\al x} $ \\ $ - \lm
(\varphi_{\al x} \varphi_{\al x}-1)^2]$ is the pure
scalar field action. The action $S$ is invariant under the so-called
staggered fermion (SF) symmetry group, which contains a discrete set
of O(4) flavor transformations. \\
$S$ is however not invariant under the full O(4) flavor group:
There are two operators with dimension four which are generated by
the quantum fluctuations and which are invariant under
the SF symmetry group, but break O(4),
$O^{(1)}=  \sum_{x \mu} \varphi_{\mu  x}^4$,
$O^{(2)}=  \frac{1}{2} \sum_{x \mu}
(\varphi_{\mu , x+\hmu} -\varphi_{\mu  x})^2$.
In order to recover the full O(4) symmetry one has in principle to add
these operators as counterterms to the
action $S \ra S + \vep_0 O^{(1)} + \dl_0 O^{(2)}$
and tune their coefficients $\vep_0$ and $\dl_0$ as a function
of the bare parameters such that the O(4) invariance
gets restored in the scaling region. We showed however in ref.~[4]
by using renormalized perturbation theory
that the deviations in the renormalized couplings due
to the O(4) symmetry breaking (SB) are very small.
The results  presented  in the following section
have been  obtained without adding these counter terms to  the action. \\
Since we are interested in the largest possible renormalized couplings
we have fixed in the numerical simulation $\lm=\infty$.
For the use of the Hybrid Monte Carlo algorithm
it is furthermore necessary to use two mass-degenerate doublets.
The $\kp$-$y$ phase diagram is shown in fig.~1.
There are four different phases, a paramagnetic (PM), a broken
or ferromagnetic (FM), an antiferromagnetic (AM) and a ferrimagnetic (FI)
phase. The various symbols mark the points in the FM phase where we
carried out numerical simulations on lattices ranging in size
from $6^3 24$ to $16^3 24$.
\maketitle
\vskip 5mm
{\bf \noindent Results of the Simulation}
\vskip 2.5mm
The renormalized couplings have been computed from the usual
tree level relations $y_R=m_F/v_R$ and $\lm_R=m_{\sg}^2/2v_R^2$,
where the renormalized field expectation value is defined as
$v_R=v/\sqrt{ Z_{\pi} }$. Here $v$ is the unrenormalized scalar
field expectation value and $Z_{\pi}$ is
the wave-function renormalization constant of the Goldstone propagator.
For the determination of the quantities $m_F$, $m_{\sg}$ and $Z_{\pi}$
we have measured the fermion-, $\sg$-particle and Goldstone-propagators
in momentum space. The fermion propagator could be well described for all
momenta by a one pole ansatz, which is characteristic for
weakly interacting fermions. The numerical results for the $\sg$-particle
and Goldstone-propagators have been fitted to the ansatz
$G_{\sg,\pi}^{-1}(p)=(\phat + m_{\sg,\pi}^2+ \Sigma_{sub}(p))/Z_{\sg,\pi}$,
where $\Sg_{sub}$ is the subtracted one fermion loop self-energy and
$\phat$ is the lattice momentum squared.
We kept $m_{\pi}$ as a fit-parameter, since the O(4) SB                effects
give rise to a small non-zero value for $m_{\pi}$, which however is much
below the other masses. The  inclusion of the self-energy
turns out to be essential if one wants to obtain reliable results
for $m_{\sg,\pi}$ and $Z_{\sg,\pi}$ also on smaller volumes.
The fact that
the one-loop ansatz is sufficient to describe the numerical results
perfectly over a large momentum interval
indicates already that the renormalized couplings are small.  \\
As a next step we have to extrapolate the finite volume results
for $y_R$ and $\lm_R$ to the infinite volume. We carried out
the simulations on lattices of size $L^3 24$ with $L$ ranging from $6$ to
$16$. For the extrapolation to the infinite volume we use a $1/L^2$  ansatz
which is expected to hold, if the spectrum contains
massless Goldstone bosons.  In our case the Goldstone bosons in the
infinite volume limit acquire a small mass
due to the O(4) SB. The fact that we have not
observed significant deviations  from the $1/L^2$ behavior
gives further evidence that the SB               effects are small. \\
In the lower graph of fig.~1 we display the infinite volume results
for the ratios $m_{\sg}/v_R=\sqrt{2\lm_R}$ and $m_F/v_R=y_R$. The symbols
in the upper and lower diagrams of fig.~1 match,
so that one can see
where in the phase diagram the results for the ratios have been
obtained. It can be seen that the numerical values for neither
ratio change when lowering $\kp$ beyond
$\kp=0$.  All points have roughly the same value of the cut-off in
units of the scalar field expectation value $v_R \approx 0.15-0.25$.
The arrows in fig.~1 mark the
tree level unitarity bounds for $\lm_R$
and $y_R$. The graph shows that
the points obtained in the regions (II) and (III)
of the phase diagram (see fig.~1) are still very close to these values,
which indicates that the renormalized couplings are not very strong.
The solid line encloses the allowed regions obtained
by integrating numerically the one-loop $\beta$-functions from
the cut-off scale down to the renormalization scale
and identifying the couplings
at these scales with the bare and the renormalized  couplings.
The cut-off was adjusted such that the agreement with
the numerical data is best. It is remarkable that the
shape is in reasonable agreement with our data.
Fig.~1 shows that the Yukawa interaction gives only a modest
increase in $\lm_R$. All in all we conclude that the renormalized
quartic and Yukawa couplings are in accordance with
triviality and that they cannot be strong, unless
the cut-off is unacceptably low.\\

The numerical calculations were performed on the CRAY Y-MP4/464
at SARA, Amsterdam, on the S600 at RWTH Aachen
and on the CRAY Y-MP/832 at HLRZ J\"ulich.
This research was supported by the ``Stich\-ting voor
Fun\-da\-men\-teel On\-der\-zoek der Materie (FOM)'',
by the ``Stichting Nationale Computer Faciliteiten (NCF)'' and by
the DOE under contract DE-FG03-90ER40546.
\maketitle
\vskip 4mm
{\bf \noindent References}
\vskip 2.0mm
\noindent [1]     I.~Montvay, {\it Phys. Lett. B}{\bf 199} (1987) 89.\\
\noindent [2]     J. Smit, {\it Nucl. Phys. B (Proc. suppl.)} {\bf 4} (1988)
451;
                 {\it Nucl. Phys. B (Proc. Suppl.)} {\bf 26} (1992)
                 480.\\
\noindent [3]    W.~Bock, J.~Smit and J.C.~Vink, {\it Phys. Lett. B}{\bf 291}
(1992)
                 297. \\
\noindent [4]    W.~Bock, C.~Frick, J.~Smit and J.C.~Vink, {\it Nucl. Phys. B
                 }{\bf 400} (1993) 309 ; {\it
                 Nucl. Phys. B (Proc. Suppl.)} {\bf 30} (1993) 643.\\
\end{document}